\documentclass[aps,prb,reprint,showpacs,superscriptaddress]{revtex4-1}
\usepackage{amsmath}
\usepackage{graphicx}
\usepackage{bm}
\usepackage{mathptmx}
\begin{document}
\title{Specular Andreev reflection in inversion-symmetric Weyl-semimetals}
\author{Wei Chen}
\affiliation{National Laboratory of Solid State Microstructures and Department of Physics, Nanjing University, Nanjing 210093, China}
\author{Liang Jiang}
\affiliation{National Laboratory of Solid State Microstructures and Department of Physics, Nanjing University, Nanjing 210093, China}
\author{R. Shen}
\affiliation{National Laboratory of Solid State Microstructures and Department of Physics, Nanjing University, Nanjing 210093, China}
\email{shen@nju.edu.cn}
\author{L. Sheng}
\affiliation{National Laboratory of Solid State Microstructures and Department of Physics, Nanjing University, Nanjing 210093, China}
\author{B. G. Wang}
\affiliation{National Laboratory of Solid State Microstructures and Department of Physics, Nanjing University, Nanjing 210093, China}
\affiliation{National Center of Microstructures and Quantum Manipulation, Nanjing University, Nanjing 210093, China}
\author{D. Y. Xing}
\affiliation{National Laboratory of Solid State Microstructures and Department of Physics, Nanjing University, Nanjing 210093, China}
\begin{abstract}
The electron-hole conversion at the normal-metal superconductor interface in inversion-symmetric Weyl semimetals is investigated with an effective two-band model. We find that the specular Andreev reflection of Weyl fermions has two unusual features. The Andreev conductance for $s$-wave BCS pairing states is anisotropic, depending on the angle between the line connecting a pair of Weyl points and the normal of the junction, due to opposite chirality carried by the paired electrons. For the Fulde-Ferrell-Larkin-Ovchinnikov pairing states, the Andreev reflection spectrum is isotropic and is independent of the finite momentum of the Cooper pairs.
\end{abstract}
\pacs{74.45.+c,73.23.-b}
\maketitle
In recent years, a great progress in condensed matter physics has been made by the discovery of topological insulators \cite{Kane}, which have a bulk energy gap and gapless surface states protected by the time reversal symmetry. Such a phase is characterized by the topological invariants rather than any order parameters. Currently, the topological matter is further extended to the Weyl semimetals (WSMs) \cite{Wan}, where the bulk energy band is gapless and the Weyl points (WPs) emerge separately in the momentum space in pairs carrying opposite chirality. Electrons around the WPs can be well described by the two-component Weyl equations. In order to realize this fascinating phase, either the time-reversal symmetry \cite{Wan, Ran, Balents, Cho} or the inversion symmetry \cite{Murakami} needs to be broken. The WSM phase is protected by the band topology and therefore is robust. The Dirac type dispersion cannot be gapped unless there exists coupling between the pair of WPs of opposite chirality \cite{Wan, Ran}.

When the superconducting pair potential is introduced into those topological matters through proximity effects \cite{Fu1} or intrinsic phonon-mediated attractive interaction \cite{Fu2}, a superconducting phase with non-trivial topology can emerge. Taking the topological insulator as an example, the $s$-wave pair potential can result in a topological superconductor with Majorana surface states \cite{Kane, Fu1, Fu2}. Recently, such a route has been adopted by Meng \textit{et al.} \cite{Meng} and Cho \textit{et al.} \cite{Moore} in the study of superconductivity in inversion-symmetric WSMs. They found a novel BCS pairing state which is an electronic analogue of $^{3}$He-A phase and the Majorana modes can emerge under certain conditions \cite{Meng, Moore}. Apart from the BCS pairing states, the separation of WPs in momentum space also provides an opportunity for the Fulde-Ferrell-Larkin-Ovchinnikov (FFLO) pairing states \cite{Fulde, Larkin}, where the Cooper pairs carry a finite momentum. In contrast to the conventional case, the FFLO state in WSMs is robust against the weak disorder \cite{Moore}.

The study on the superconducting WSMs is still in the preliminary stage and one central question to be resolved is how to detect them. One of the effective methods for detecting pairing potentials is through the Andreev reflection (AR) \cite{Andreev} at the normal-metal superconductor (NS) interface. The AR spectrum has been successfully used to probe both the conventional $s$-wave \cite{BTK} and the unconventional anisotropic \cite{Bruder, Tanaka, Yamashiro} superconductors. The recent progress of the studies on the AR also reveals some unusual features, such as the specular AR in the graphene \cite{Beenakker} and the resonant AR induced by the Majorana modes \cite{Nagaosa, Law}.

\begin{figure}
\centering
\includegraphics[width=0.45\textwidth]{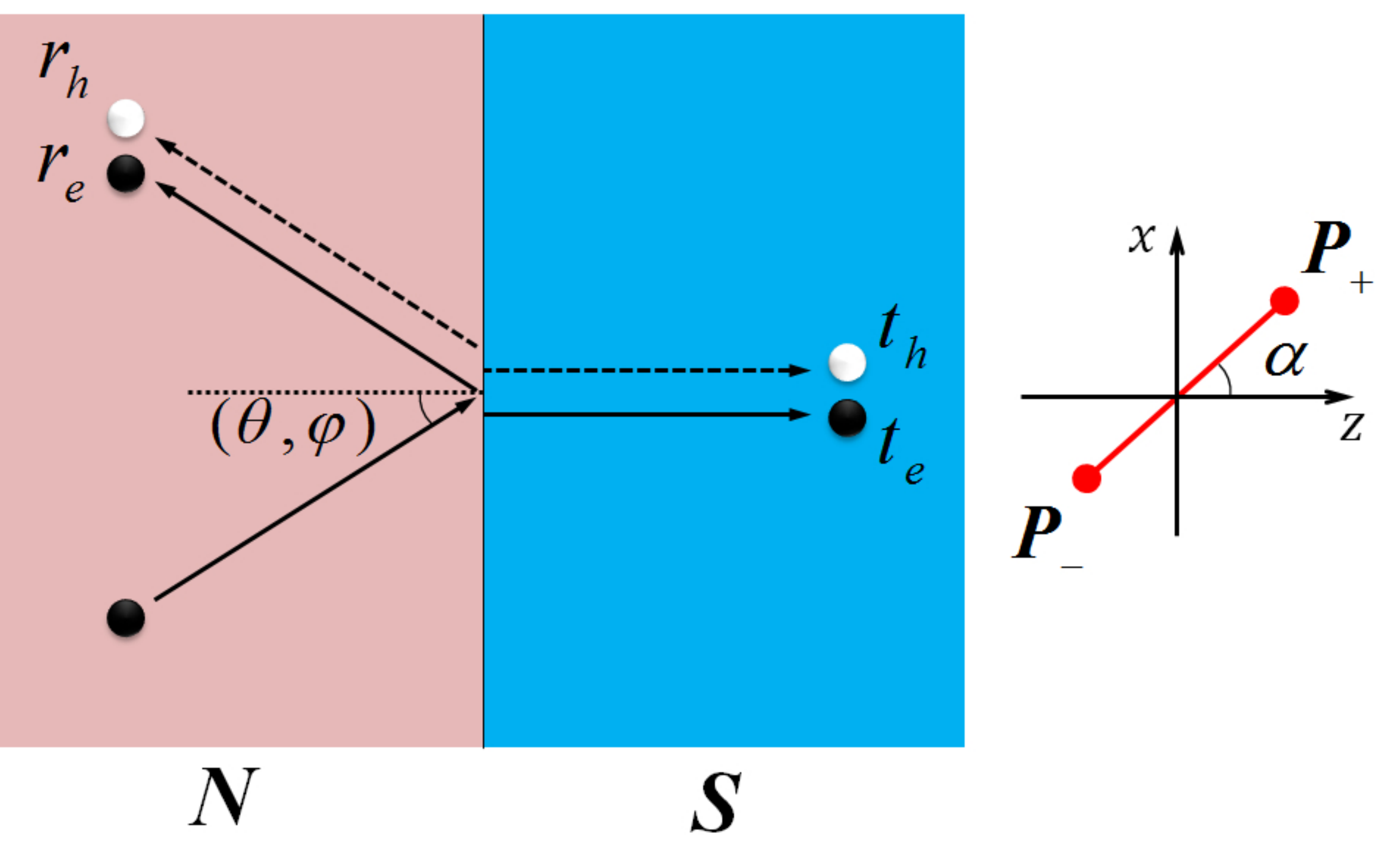}
\caption{(Colour on-line) Illustration of the NS junction composed of WSMs. The black and white balls represent the electrons and holes, respectively. An electron of incident angle $(\theta ,\varphi)$ from the normal-metal side (the left) is scattered into four branches: the normal reflection, the specular AR, the electron-like quasiparticle transmission, and the hole-like quasiparticle transmission, with amplitudes $r_{e}$, $r_{h}$, $t_{e}$, and $t_{h}$, respectively. Line $\bm{P}_{-}\bm{P}_{+}$ makes an angle $\alpha$ with the normal of the interface.}\label{fig1}
\end{figure}

In this paper, we employ an effective two-band model to investigate the specular AR at the NS interface in inversion-symmetric WSMs with either the BCS or the FFLO pairing states. It is found that the normalized subgap conductance for the $s$-wave BCS pairing states is a constant of 1.726, and the effective gap is anisotropic, which is a sinusoidal function of the angle between the axis connecting two WPs and the normal of the junction. For the FFLO pairing states, the AR spectrum exhibits an unexpected isotropic behavior, being nothing to do with the finite momentum of the Cooper pairs.

The NS junction under consideration is sketched in fig. \ref{fig1}. The normal of the junction is along the $z$-axis and the interface is located at $z=0$. The superconducting WSM can be prepared by doping \cite{Moore}, which drops the WPs well below the Fermi level. The doping level should be chosen properly, so that the electrons at the Fermi level have a large density of states for the phonon-mediated condensation but they can still be described by the Weyl equation. The normal metal in fig. \ref{fig1} is a normal WSM without any doping.

The minimal model for inversion-symmetric WSMs is described by an effective two-band Hamiltonian (in units of $\hbar=1$) \cite{Moore} 
\begin{equation}\label{H0}
\begin{split}
&H_{0}=\sum_{a=\pm}c_{a}^{\dagger}(\bm{q})h_{a}(\hat{R}^{-1}\bm{q},\hat{R}^{-1}\bm{\sigma})c_{a}(\bm{q}),\\
&h_{\pm}(\bm{q},\bm{\sigma})=v_{0}(q_{x}\sigma_{x}+q_{y}\sigma_{y}\mp q_{z}\sigma_{z})-U\Theta(z),
\end{split}
\end{equation}
where $c_{a}=(c_{a\uparrow},c_{a\downarrow})^{T}$ is the electron operator with $a=\pm$ denoting two Weyl nodes, respectively, $v_{0}$ is the Fermi velocity, $\bm{q}$ is the momentum measured from the WPs, $\bm{\sigma}$ is the Pauli matrix of spins, and $\Theta(z)$ is the step function. Each Weyl node contains two cylindrical three-dimensional cones that touch at the WP. In this model, a pair of WPs are located at two inversion-symmetric points, $\bm{P}_{+}=\bm{P}$ and $\bm{P}_{-}=-\bm{P}$. In general, the axis connecting two WPs, line $\bm{P}_{-}\bm{P}_{+}$, can make an angle of $\alpha$ with the normal of the junction, as shown in fig. \ref{fig1}. Therefore, the rotation operator $\hat{R}^{-1}=\begin{pmatrix}\cos\alpha & -\sin\alpha \\ \sin\alpha & \cos\alpha\end{pmatrix}$ is incorporated in eq. (\ref{H0}), with which a vector is rotated around the $y$-axis by an angle $\alpha$. When $\alpha =0$, line $\bm{P}_{-}\bm{P}_{+}$ is along the $z$-axis and the Hamiltonian in Ref. \cite{Moore} is recovered. The energy in eq. (\ref{H0}) is measured from the Fermi level. In the normal-metal side ($z<0$), there is no band shift and the WPs are at the Fermi level. In the superconductor side ($z>0$), there is a large band shift $U$ caused by doping, so that the WPs are well below the Fermi level. Although the effective Hamiltonian (\ref{H0}) contains only one pair of WPs, it can well describe the low energy excitations of two proposed candidates for WSMs based on topological insulators \cite{Balents, Cho}.

The electrons in WSMs bear the spin-momentum locking features and nodes $\bm{P}_{\pm}$ carry opposite chirality. As a result, there are two types of pairing mechanisms \cite{Moore}: one is the inter-node BCS pairing for which the paired electrons are from two Weyl nodes of opposite chirality and the momentum of the Cooper pair is zero, the other is the intra-node FFLO pairing for which the paired electrons are from only one Weyl node and the momentum of the Cooper pair is finite ($2\bm{P}_{\pm}$). The Hamiltonian for both pairing states can be expressed as
\begin{equation}\label{H1}
\begin{split}
&H^{B}_{\text{pair}}=\sum_{\bm{q},a}\Delta(z) c^{\dagger}_{a\uparrow}(\bm{q})c^{\dagger}_{-a\downarrow}(-\bm{q})+h.c.,\\
&H^{F}_{\text{pair}}=\sum_{\bm{q},a}\Delta(z) c^{\dagger}_{a\uparrow}(\bm{q})c^{\dagger}_{a\downarrow}(-\bm{q})+h.c.,
\end{split}
\end{equation}
where superscripts $B$ and $F$ correspond to the BCS and FFLO pairing states, respectively, and $\Delta(z)=\Delta\Theta(z)$ is the $s$-wave pairing potential, which is chosen to be real.

After the diagonalization of the Hamiltonian $H=H_{0}+H^{B(F)}_{\text{pair}}$, the energy excitations are given by 
\begin{equation}\label{E}
\begin{split}
&E^{B}=\sqrt{\big(v_{0}|\bm{q}|\pm\sqrt{U^{2}+\Delta^{2}\cos^{2}\beta}\big)^{2}+\Delta^{2}\sin^{2}\beta},\\
&E^{F}=\sqrt{(v_{0}|\bm{q}|\pm U)^{2}+\Delta^{2}},
\end{split}
\end{equation}
where $\beta$ is the azimuthal angle of the momentum relative to line $\bm{P}_{-}\bm{P}_{+}$. Due to the different manners of spin-momentum locking, the electrons from two Weyl nodes cannot fully paired \cite{Moore}. The excitation spectrum $E^{B}$ is consequently characterized by an anisotropic effective gap $|\Delta\sin\beta|$. The nodes of $E^{B}$ exactly lie in line $\bm{P}_{-}\bm{P}_{+}$ ($\beta=0,\pi$). On the contrary, the Cooper pairs in FFLO superconductors are formed within the same Weyl node and the excitation spectrum $E^{F}$ is fully gapped. 

The Bogoliubov-de Gennes (BdG) Hamiltonian in the real space can be obtained by performing a Fourier transformation on eqs. (\ref{H0}) and (\ref{H1}). In the Nambu representation where $\hat{\psi}=[\psi_{\uparrow}(\bm{r}),\psi_{\downarrow}(\bm{r}),\psi^{\dagger}_{\downarrow}(\bm{r}),-\psi^{\dagger}_{\uparrow}(\bm{r})]^{T}$, the BdG Hamiltonians for both pairing cases are given by
\begin{equation}\label{bdg}
\begin{split}
&H^{B}_{\text{BdG}}=\begin{pmatrix}
h_{\pm}'(-i\bm{\nabla}-\bm{P}_{\pm},\bm{\sigma}) & \Delta \Theta (z)\\
\Delta \Theta (z) & -h_{\mp}'(-i\bm{\nabla}-\bm{P}_{\pm},\bm{\sigma})
\end{pmatrix},\\
&H^{F}_{\text{BdG}}=\begin{pmatrix}
h_{\pm}'(-i\bm{\nabla}-\bm{P}_{\pm},\bm{\sigma}) & \Delta e^{2i\bm{P}_{\pm}\cdot\bm{r}} \Theta (z)\\
\Delta e^{-2i\bm{P}_{\pm}\cdot\bm{r}} \Theta (z) & -h_{\pm}'(-i\bm{\nabla}-\bm{P}_{\mp},\bm{\sigma})
\end{pmatrix},
\end{split}
\end{equation}
where $h'_{\pm}(\bm{\nabla},\bm{\sigma})=h_{\pm}(\hat{R}^{-1}\bm{\nabla},\hat{R}^{-1}\bm{\sigma})$. The pairing potential exhibits a spatial modulation for the FFLO superconductors and is a constant for the BCS ones.

Considering an electron of energy $E$ and of incident angle $(\theta, \varphi)$ coming from the normal-metal side, it can be normally reflected as an electron or Andreev reflected as a hole, or transmitted into the superconductor side as the electron-like or hole-like quasiparticles, as shown in fig. \ref{fig1}. As the Fermi level is located exactly at the WPs in the normal-metal side, the incident electron and the reflected hole must belong to the conduction and the valence band, respectively, resulting in the specular AR, which is similar to what occurs in the graphene \cite{Beenakker}.

For the BCS pairing WSM, an incident electron in node $\bm{P}_{+}$ can be converted into a hole in node $\bm{P}_{-}$ via the AR process. The momenta of the electron and hole are both around $\bm{P}_{+}$, so that the BdG equation can be solved with the ansatz $\psi =(fe^{i\bm{P}_{+}\cdot\bm{r}},ge^{i\bm{P}_{+}\cdot\bm{r}})^{T}$, where the $f$ and $g$ are the electron and hole components of the wave function. For the FFLO pairing WSM, an incident electron in node $\bm{P}_{-}$ is converted into a hole still in node $\bm{P}_{-}$. The momenta of the electron and hole are around $\bm{P}_{-}$ and $\bm{P}_{+}$, respectively, and the BdG equation can be solved with the ansatz $\psi =(fe^{i\bm{P}_{-}\cdot\bm{r}},ge^{i\bm{P}_{+}\cdot\bm{r}})^{T}$.

For the BCS pairing case, the wave function in the normal-metal side $\Psi_{N}^{B}$ and that in the superconductor side $\Psi_{S}^{B}$ are given by
\begin{equation}\label{BCS_wave}
\begin{split}
&\Psi_{N}^{B}=\hat{b}_{1}e^{iq\cos\theta z}+r_{e}^{B}\hat{b}_{2}e^{-iq\cos\theta z}+r_{h}^{B}\hat{b}_{3}e^{-iq\cos\theta z},\\
&\Psi_{S}^{B}=t_{e}^{B}\hat{b}_{4}e^{iq_{z+}^{B}z}+t_{h}^{B}\hat{b}_{5}e^{iq_{z-}^{B}z},
\end{split}
\end{equation}
where $\hat{b}_{1}=[s(\alpha,\theta), s(\alpha-\pi /2,\theta), 0, 0]^{T}$, $\hat{b}_{2}=\hat{b}_{1}(\pi -\theta)$, $\hat{b}_{3}=[0, 0, e^{-i\varphi /2}\cos\theta /2,-e^{i\varphi /2}\sin\theta /2]^{T}$, $\hat{b}_{4}=[\sin 2\alpha, -2\cos^{2}\alpha, \Delta\sin 2\alpha /(E+E\gamma_{B}),0]^{T}$, and $\hat{b}_{5}=[\sin 2\alpha, 2\sin^{2}\alpha, 0, 2\Delta\sin^{2}\alpha /(E-E\gamma_{B})]^{T}$ are the basis functions for the incident electron, the normal reflected electron, the Andreev reflected hole, and the electron-like and hole-like transmitted quasiparticles, respectively, with $s(\alpha,\theta)=\cos\alpha\sin(\theta /2)e^{-i\varphi /2}-\sin\alpha\cos(\theta /2)e^{i\varphi /2}$, and $\gamma_{B}=\sqrt{E^{2}-\Delta^{2}\sin^{2}\alpha}/E$. The scattering coefficients for each process are denoted by $r_{e,h}^{B}$ and $t_{e,h}^{B}$, respectively. The wave vectors are obtained as $q=E/v_{0}$ in the normal-metal side, and $q_{z\pm}^{B}=(\pm U+E\gamma_{B})/v_{0}$ for the electron-like and hole-like quasiparticles in the superconductor side. The rapid oscillation factor $e^{i\bm{P}_{+}\cdot\bm{r}}$ in the ansatz and the transverse plane wave $e^{i\bm{q}_{\parallel}\cdot\bm{r}}$ are omitted in eq. (\ref{BCS_wave}) for simplicity, as the transverse momentum $\bm{q}_{\parallel}$ is conserved during the scattering. In eq. (\ref{BCS_wave}), the heavy doping condition $U\gg E, \Delta$ is also utilized. As a result, there is a large mismatch of wave vectors between two sides, and the quasiparticles in the superconductor side are nearly perpendicularly transmitted.

For the FFLO pairing case, the wave function can be obtained in a parallel manner, and reads
\begin{equation}\label{FFLO_wave}
\begin{split}
&\Psi_{N}^{F}=\hat{f}_{1}e^{iq\cos\theta z}+r_{e}^{F}\hat{f}_{2}e^{-iq\cos\theta z}+r_{h}^{F}\hat{f}_{3}e^{-iq\cos\theta z},\\
&\Psi_{S}^{F}=t_{e}^{F}\hat{f}_{4}e^{iq_{z+}^{F}z}+t_{h}^{F}\hat{f}_{5}e^{iq_{z-}^{F}z},
\end{split}
\end{equation}
where the basis functions are written as $\hat{f}_{1}=[(e^{-i\varphi /2}\cos\theta /2,e^{i\varphi /2}\sin\theta /2)e^{i\bm{P}_{-}\cdot\bm{r}}, 0, 0]^{T}$, $\hat{f}_{2}=\hat{f}_{1}(\pi-\theta)$, $\hat{f}_{3}=[0, 0, (e^{-i\varphi /2}\cos\theta /2 ,-e^{i\varphi /2}\sin\theta /2)e^{i\bm{P}_{+}\cdot\bm{r}}]^{T}$, $\hat{f}_{4}=[e^{i\bm{P}_{-}\cdot\bm{r}}, 0, e^{i\bm{P}_{+}\cdot\bm{r}}E(1-\gamma_{F})/\Delta,0]^{T}$, and $\hat{f}_{5}=[0,e^{i\bm{P}_{-}\cdot\bm{r}},0,e^{i\bm{P}_{+}\cdot\bm{r}}E(1+\gamma_{F})/\Delta]^{T}$, respectively, with $\gamma_{F}=\sqrt{E^{2}-\Delta^{2}}/E$ and the wave vectors $q_{z\pm}^{F}=(\pm U+E\gamma_{F})/v_{0}$. One finds that the rapid oscillation factors $e^{i\bm{P}_{\pm}\cdot\bm{r}}$ for the electron and hole components of the wave functions are different for the FFLO superconductor.

By matching the boundary condition $\Psi_{N}^{B(F)}=\Psi_{S}^{B(F)}$ at the interface $z=0$, one obtains the scattering amplitudes as
\begin{equation}\label{scattering}
\begin{split}
&r_{h}^B=-\frac{(\Delta\sin\alpha /E)\cos\theta}{\cos\varphi(\gamma_{B}+\cos\theta)-i\sin\varphi(1+\gamma_{B}\cos\theta)},\\
&r_{e}^{B}=\frac{\sin\theta(-\gamma_{B}\cos\varphi+i\sin\varphi)}{\cos\varphi(\gamma_{B}+\cos\theta)-i\sin\varphi(1+\gamma_{B}\cos\theta)},\\
&r_{h}^{F}=\frac{(\Delta/E)\cos\theta}{1+\gamma_{F}\cos\theta},\\
&r_{e}^{F}=-\frac{\sin\theta}{1+\gamma_{F}\cos\theta}.
\end{split}
\end{equation}

With the help of the Blonder-Tinkham-Klapwijk (BTK) formula \cite{BTK}, for a specific incident angle, the differential conductance is determined by the BTK coefficient
\begin{equation}\label{conductance_1}
S^{i}(E,\theta,\varphi)=1+|r_{h}^{i}(E,\theta,\varphi)|^{2}-|r_{e}^{i}(E,\theta,\varphi)|^{2},
\end{equation}
with $i=(B,F)$. In units of the conductance of the junction in the normal state ($\Delta =0$), the normalized conductance is given by  
\begin{equation}\label{conductance_2}
G^{i}(eV)=\frac{\int_{0}^{\pi /2}d\theta\sin2\theta\int_{0}^{2\pi}d\varphi S^{i}(eV,\theta,\varphi)}{\int_{0}^{\pi /2}d\theta\sin2\theta\int_{0}^{2\pi}d\varphi S^{i}(eV,\theta,\varphi)|_{\Delta =0}},
\end{equation}
where $eV$ is the bias voltage.

Firstly, we consider the AR spectrum for the BCS pairing states. Although the pair potential of an $s$-wave BCS pairing WSM is a constant, the electrons with different spin-momentum locking features can only partly paired according to their momentum direction, leading to an anisotropic effective gap $|\Delta\sin\beta|$ in energy excitation $E^{B}$. Such an anisotropic gap is also exhibited in the AR spectrum. Under the heavy doping condition, the quasiparticles are nearly perpendicularly propagated in the superconducting WSM. The azimuthal angle of the momentum relative to line $\bm{P}_{-}\bm{P}_{+}$ can be approximated as $\alpha$. Therefore, one finds that the AR coefficient $r^{B}_{h}$ is proportional to $\Delta\sin\alpha$ and the $\gamma_{B}$ is also determined by $\Delta\sin\alpha$. When $\alpha=0$ or $\pi$, the effective gap is zero and the AR is totally suppressed. 

\begin{figure}
\centering
\includegraphics[width=0.45\textwidth]{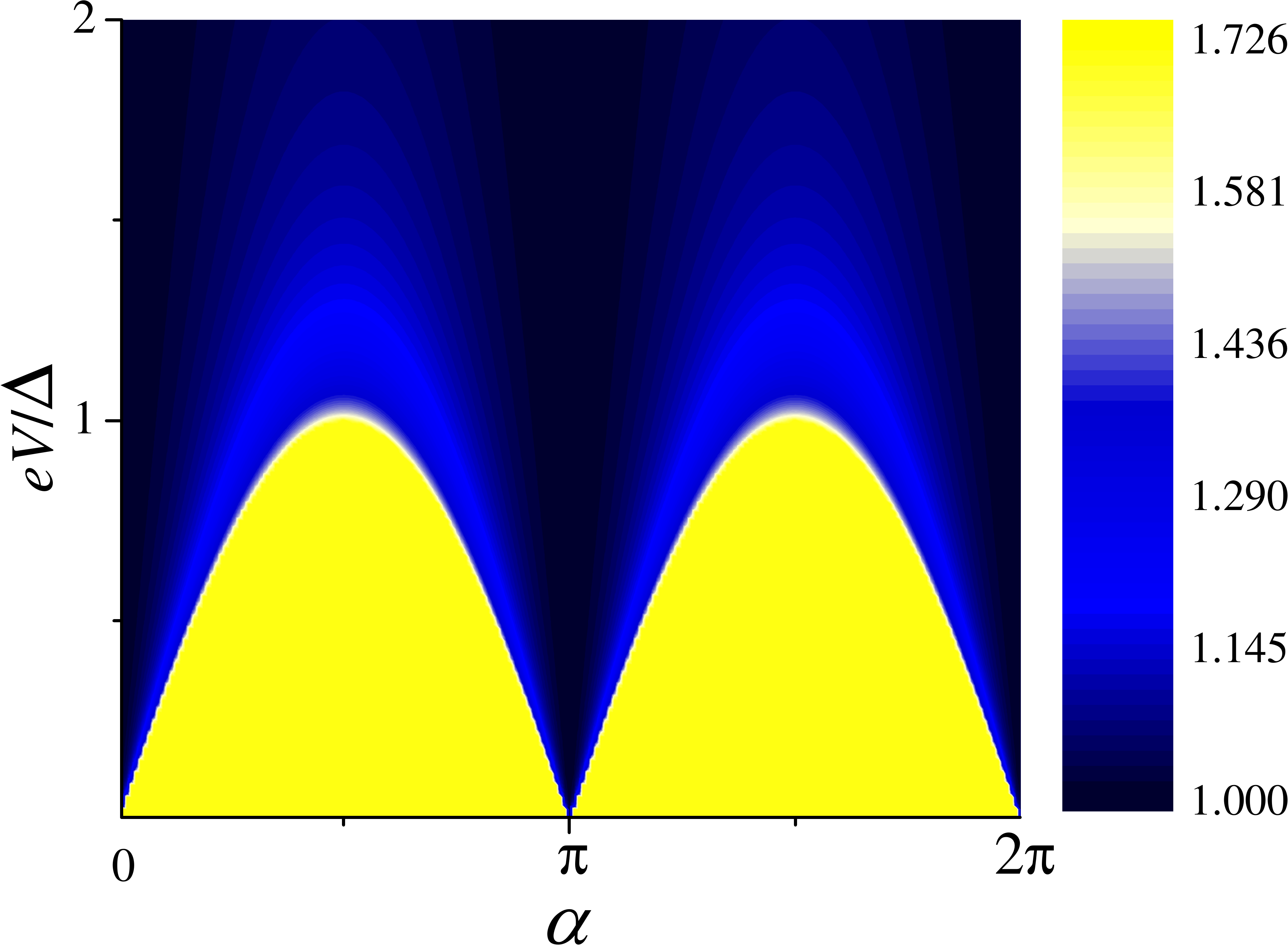}
\caption{(Colour on-line) Contour plot of the normalized differential conductance of the NS junction with the BCS pairing states, as a function of $eV$ and $\alpha$. }\label{contour}
\end{figure}

The dependence of the normalized conductance on both the bias voltage and angle $\alpha$ is plotted in fig. \ref{contour}. A notable characteristic in fig. \ref{contour} is the yellow region of a contour of $|\Delta\sin\alpha|$, in which the normalized conductance is a constant of 1.726. Such a constant behavior can be understood as follows. When the incident electron is of an energy below the effective gap, the transmitted wave is evanescent and $\gamma_{B}$ is pure imaginary, so that the subgap conductance is determined by the BTK coefficient
\begin{equation}\label{within_gap}
S^{B}_{<}=2|r_{h}^{B}|^2=\frac{2\cos^{2}\theta}{\cos^{2}\theta\cos^{2}(\varphi-\eta)+\sin^{2}(\varphi-\eta)},
\end{equation}
with $\eta=\arctan(\sqrt{\Delta^{2}\sin^{2}\alpha-E^{2}}/E)$. For the perpendicular incidence ($\theta =0$), the normal reflection of Weyl fermions is completely suppressed due to the spin-momentum locking, and BTK coefficient $S^{B}_{<}$ is a constant of 2, as in the conventional NS junction in the transparent limit \cite{BTK}. For the oblique incidence, there is a finite normal reflection caused by the large momentum mismatch at the interface, and $S^{B}_{<}$ in eq. (\ref{within_gap}) is less than 2 and depends on the energy in general. Interestingly, the energy factor in $S^{B}_{<}$ is always bound with the incident angle $\varphi$. Therefore, after an integration over incident angles and normalization, the subgap conductance becomes a constant of 1.726, independent of energy. For the same reason, the subgap noise power of the junction is also a constant.

\begin{figure}
\centering
\includegraphics[width=0.45\textwidth]{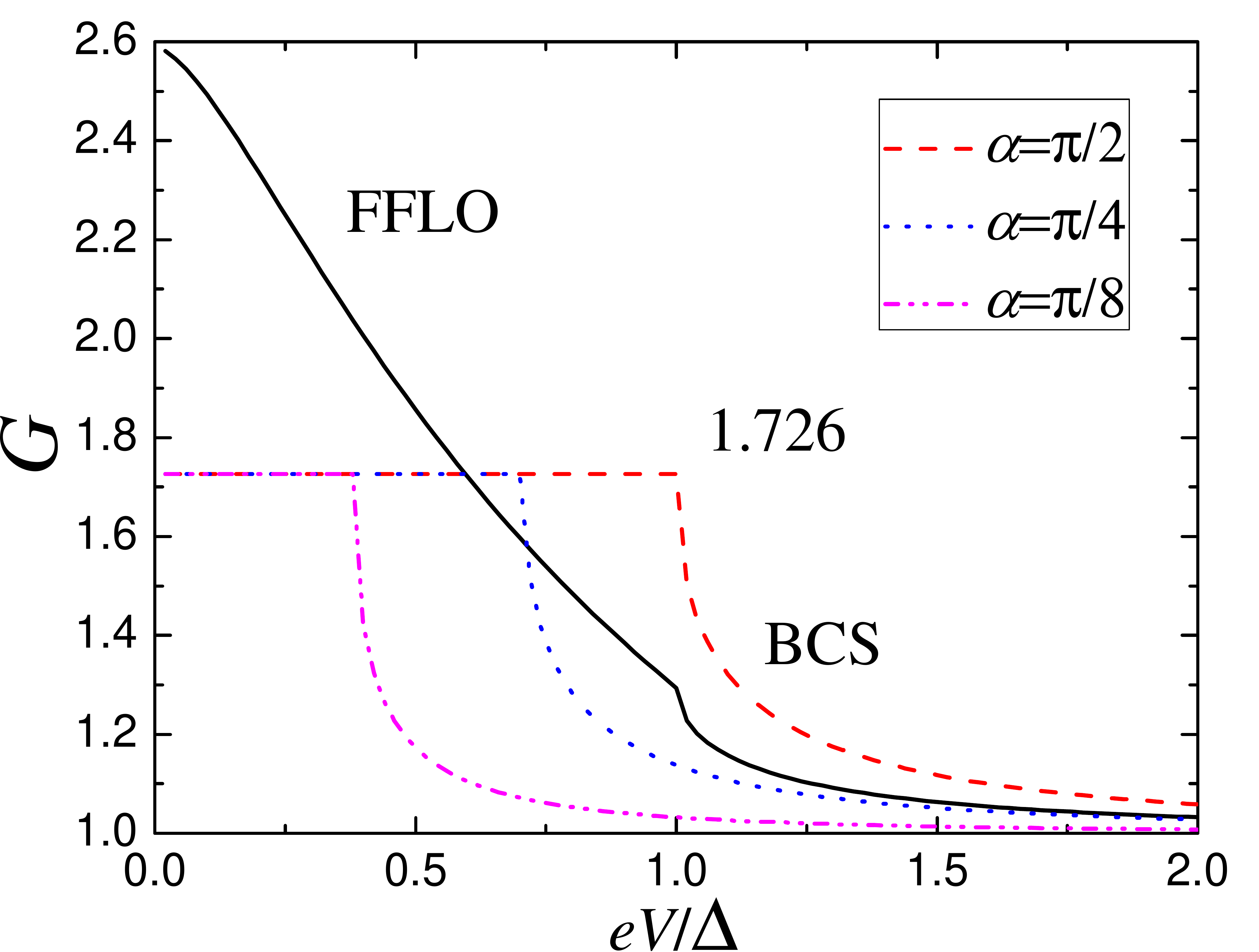}
\caption{(Colour on-line) Plots of the normalized differential conductance as a function of $eV$. The solid line is corresponding to the FFLO pairing states. The dashed, dotted and dot-dashed lines are corresponding to the BCS pairing states with different $\alpha$, respectively.}\label{results}
\end{figure}

Secondly, we consider the AR spectrum for the FFLO pairing states, which is plotted in fig. \ref{results} as the solid line. Although the Cooper pair in an FFLO pairing WSM bears a finite momentum of $2\bm{P}_{\pm}$ and the pair potential is modulated in the spatial space, the AR spectrum is still isotropic and has nothing to do with $\bm{P}_{\pm}$. This is due to three reasons: a) In contrast to the conventional FFLO superconductor, the momentum of the Cooper pair in WSMs arises from the finite momentum of the WP and does not affect the energies of two paired electrons. b) The Weyl fermions are linearly dispersed and therefore their velocities are independent of their momenta. In order to guarantee the current conservation, only the continuity of the wave function needs to be considered while the first derivative of the wave function does not help, so that momentum $\bm{P}_{\mp}$ carried by electrons and holes plays no role in the AR spectrum. c) The FFLO pairing states are formed within one Weyl node and thus the AR spectrum should also be independent of $\alpha$, which is determined by a line connecting two WPs.  Since the energy excitation is fully gapped and neither $\bm{P}_{\pm}$ nor $\alpha$ makes contribution, the AR spectrum of an FFLO pairing WSM is isotropic and resembles that of a BCS pairing graphene \cite{Beenakker}. 

The normalized AR spectrum for the BCS pairing WSMs is also plotted in fig. \ref{results} for comparison, which is a constant of 1.726 below the anisotropic effective gap $|\Delta\sin\alpha|$. It is clearly that the BCS and the FFLO pairing states are effectively distinguished by their AR spectra.

It is helpful to discuss the experimental realization of the NS junction in the WSMs. The present effective model of the WSM is actually corresponding to the topological insulator multilayers \cite{Balents} and magnetically doped Bi$_{2}$Se$_{3}$ \cite{Cho}. Several recent experimental progress on topological insulators, such as the control of the mass term in BiTl(S$_{1-\delta}$Se$_{\delta}$)$_{2}$ \cite{Xu}, the simultaneous magnetic and charge doping \cite{Chen}, and the growth of the ultrathin high-quality films of Bi$_{2}$Se$_{3}$ \cite{Zhang}, paves a way to the realization of the WSM in topological insulator systems. The recent realization of superconductivity in the doped topological insulators \cite{Wray} also indicates a hopeful future for the superconducting WSMs.

In summary, the specular AR in the NS junction in inversion-symmetric WSMs is investigated. The AR spectrum for the BCS pairing states indicates an anisotropic effective gap $|\Delta\sin\alpha |$, below which the normalized conductance is a constant of 1.726. The AR spectrum for the FFLO pairing states is independent of the momentum of the Cooper pair and is isotropic. The BCS and FFLO pairing states can be easily distinguished by their AR spectra. 

\begin{acknowledgments}
We would like to thank X. Wan, Jing-Min Hou and L. B. Shao for helpful discussions. This work was supported by 973 Program (Grants No. 2011CB922103, No. 2011CBA00205, and No. 2009CB929504), by NSFC (Grants No. 11074111, No. 11174125, and No. 11023002), by PAPD of Jiangsu Higher Education Institutions, by NCET, and by the Fundamental Research Funds for the Central Universities.
\end{acknowledgments}

\end{document}